\documentclass{IEEEtran}
\IEEEoverridecommandlockouts

\usepackage{pifont}
\usepackage{cite}
\usepackage{amsmath,amssymb,amsfonts}
\usepackage{algorithmicx}
\usepackage{algpseudocode}
\usepackage{algorithm}
\usepackage{graphicx}
\usepackage{textcomp}
\usepackage{xcolor}
\usepackage{multirow}
\usepackage{enumitem}
\usepackage{tabularx}
\usepackage[normalem]{ulem}
\usepackage[left=0.63in,right=0.63in,top=0.72in,bottom=1.65in]{geometry}
\usepackage{catchfile}

\usepackage{tikz}
\usepackage{hyperref}

\def\BibTeX{{\rm B\kern-.05em{\sc i\kern-.025em b}\kern-.08em
    T\kern-.1667em\lower.7ex\hbox{E}\kern-.125emX}}

\definecolor{lime}{HTML}{A6CE39}
\DeclareRobustCommand{\orcidicon}{%
    \begin{tikzpicture}
        \draw[lime, fill=lime] (0,0) circle [radius=0.16] node[white] {\tiny ID};
        \draw[white, fill=white] (-0.0625,0.095) circle [radius=0.007];
    \end{tikzpicture}%
    \hspace{-2mm}%
}
\foreach \x in {A, ..., Z}{%
    \expandafter\xdef\csname orcid\x\endcsname{%
        \noexpand\href{https://orcid.org/\csname orcidauthor\x\endcsname}{\noexpand\orcidicon}}%
}

\usepackage{eso-pic}

\AddToShipoutPictureFG{%
    \AtPageUpperLeft{%
        \put(\LenToUnit{0.5\paperwidth},\LenToUnit{-3mm}){%
            \makebox[0pt][c]{%
                \parbox[t]{0.80\paperwidth}{%
                    \centering
                    \normalfont\sffamily\fontsize{6}{9}\selectfont
                    This article has been accepted for publication in 
                    IEEE Communications Standards Magazine. This is the
                    author's version which has not been fully edited 
                    and\\ content may change prior to final publication.
                    Citation information: DOI 10.1109/MCOMSTD.2026.3748641
                }%
            }%
        }%
    }%
}

\AddToShipoutPictureFG{%
    \AtPageLowerLeft{%
        \put(\LenToUnit{0.5\paperwidth},\LenToUnit{3mm}){%
            \makebox[0pt][c]{%
                \parbox[b]{0.80\paperwidth}{%
                    \centering
                    \normalfont\sffamily\fontsize{6}{9}\selectfont
                    \textcopyright{} 2026 IEEE. All rights reserved,
                    including rights for text and data mining and
                    training of artificial intelligence and similar
                    technologies. Personal use is permitted, \\but
                    republication/redistribution requires IEEE permission.
                    See https://www.ieee.org/publications/rights/index.html
                    for more information.
                }%
            }%
        }%
    }%
}

\begin{document}




\title{\huge rApp/xApp Attestation:\\ A New Security Use Case for O-RAN \\
\author{
    \IEEEauthorblockN{
        Hamed Alimohammadi\IEEEauthorrefmark{1}\orcidA{}, \textit{Member, IEEE},
        Burcu Şahin\IEEEauthorrefmark{2}\orcidB{},
        Arda Akman\IEEEauthorrefmark{2}\orcidC{},\\
        Chuan Heng Foh\IEEEauthorrefmark{1}\orcidD{}, \textit{Senior Member, IEEE},
        Periklis Chatzimisios\IEEEauthorrefmark{3}\orcidE{}, \textit{Senior Member, IEEE},\\
        and Mohammad Shojafar\IEEEauthorrefmark{1}\orcidF{}, \textit{Senior Member, IEEE}
    }\\
    \IEEEauthorblockA{
        \IEEEauthorrefmark{1}6G Innovation Centre, Institute for Communication Systems (ICS), University of Surrey, Guildford, UK \\
        \IEEEauthorrefmark{2}Nokia, Sunnyvale, California, United States \\
        \IEEEauthorrefmark{3}Department of Information and Electronic Engineering, International Hellenic University, Thessaloniki, Greece, and Department of Electrical and Computer Engineering, University of New Mexico, Albuquerque, USA \\
        Email: \{h.alimohammadi,c.foh,m.shojafar\}@surrey.ac.uk, \\\{burcu.sahin,arda.akman\}@nokia.com, \\\ pchatzimisios@ihu.gr
    }
}
}

\maketitle

\begin{abstract}
The disaggregation and softwarization introduced by the Open Radio Access Network (O-RAN) architecture enable multi-vendor innovation but also expose the RAN Intelligent Controller (RIC) ecosystem to new runtime security risks. Existing O-RAN specifications define strong safeguards for onboarding, authentication, identity management, and secure communication; however, they do not provide a concrete mechanism for verifying whether deployed rApps and xApps remain in their intended, untampered state during operation.

This paper introduces rApp/xApp attestation as a RIC-native O-RAN security use case for runtime integrity verification. Rather than proposing a new cryptographic protocol, the work defines how existing integrity verification techniques can be integrated into O-RAN through attestation modules, attestation agents, RIC application interfaces, and SMO-driven policy coordination. We map the use case to relevant O-RAN Alliance working groups, identify required standardization extensions, and demonstrate feasibility through a lightweight hash-based prototype implemented on the Near-RT RIC platform. Experimental results show attestation latencies below 40 ms across multiple cryptographic hash functions, indicating that runtime attestation can be performed without disrupting time-sensitive RIC operations when appropriately scheduled. Finally, we discuss remaining technical and standardization challenges, including trusted verification, known-good runtime states, scalability, mitigation policies, and future hybrid attestation mechanisms.
\end{abstract}


\section{Introduction}\label{intro}

The Open Radio Access Network (O-RAN) disaggregates traditionally integrated RANs into standardized Radio Units (O-RUs), Distributed Units (O-DUs), and Central Units (O-CUs), with open interfaces between them. This enables multi-vendor deployment, fostering competition and innovation.  

A defining feature of O-RAN is the RAN Intelligent Controller (RIC), comprising the Non-Real-Time RIC (Non-RT RIC) and Near-Real-Time RIC (Near-RT RIC). The Non-RT RIC, within the Service Management and Orchestration (SMO), operates above one-second timescales and hosts rApps for long-term optimization and analytics, while the Near-RT RIC operates between 10~ms and 1~s and hosts xApps for time-sensitive control such as traffic steering, interference management, and mobility optimization. Through A1, rApps provide policies and guidance to near-real-time xApps.

While this architecture enables flexibility and innovation, it also expands the RIC control ecosystem by introducing open interfaces, multi-vendor software components, and third-party control applications into the RAN decision-making path. Among these new exposure points, rApps and xApps are particularly security-critical because they can influence RAN behavior at both the policy and near-real-time control layers. A compromised rApp or xApp could therefore manipulate control logic, alter operational data, or disrupt network behavior.

The O-RAN Alliance recognizes these risks and has issued specifications through Working Group 11 (WG11) for onboarding, authentication, and interface protection. However, they do not ensure application integrity after deployment. The \textit{Study on Security for Application Lifecycle Management} \cite{oran-LCM} identifies runtime integrity as a potential requirement but does not define a mechanism for enforcing it. Thus, runtime trust remains an open gap, summarized in Section~\ref{background}.

Recent research has explored zero-trust security in O-RAN, most notably ZTRAN \cite{ztran}, which implements authentication, intrusion detection, and secure slicing through dedicated xApps in the Near-RT RIC to strengthen network-level security and access control. It also highlights the need for monitoring and anomaly detection to identify suspicious behavior in network operations and control actions, thereby acknowledging that rApps/xApps may deviate from their intended operation at runtime. However, the study primarily focuses on using xApps to secure the network rather than verifying the integrity of the xApps themselves, and thus does not establish runtime trust in rApps/xApps.

In parallel, other studies have examined security challenges in O-RAN applications. For instance, \cite{OJCOMS} analyzes threats related to xApp access control and the E2 interface, highlighting vulnerabilities in control-plane interactions. Similarly, \cite{TDSC} proposes a framework for authentication, authorization, and isolation of xApps, focusing on service-level protection and secure onboarding. 
While these efforts address important aspects of O-RAN security, including access control, interface protection, and service isolation, they do not address the scenario where a compromised application behaves maliciously while operating within its permitted privileges. 

A natural candidate for addressing this missing runtime assurance is remote attestation. Techniques developed in trusted computing, cloud, and Network Functions Virtualization (NFV) environments provide mechanisms for verifying software integrity, typically through hash- or measurement-based approaches combined with challenge–response protocols. These approaches may be supported by hardware roots of trust such as Trusted Platform Modules (TPMs) and Trusted Execution Environments (TEEs), which strengthen integrity guarantees. However, such frameworks are not designed for the service-based, multi-vendor, and control-loop-driven nature of the O-RAN RIC, nor do they define how attestation integrates with RIC workflows, interfaces, and lifecycle management. Moreover, O-RAN deployments rely on heterogeneous, commercial off-the-shelf (COTS) infrastructures where uniform hardware trust anchors cannot be assumed, limiting the direct applicability of hardware-assisted approaches.

\textbf{To address this gap, this paper introduces \textit{rApp/xApp attestation} as an O-RAN \emph{security use case} for runtime integrity verification of applications operating within the RIC. }Attestation enables continuous validation of application integrity during execution, providing operators with assurance that rApps/xApps remain in their intended, untampered state. Rather than proposing a new attestation algorithm or cryptographic protocol, this work focuses on defining a RIC-native use case that integrates existing integrity verification techniques into the O-RAN architecture, aligned with its interfaces (e.g., R1 and RIC APIs), lifecycle models, and timing constraints, while identifying the standardization gaps required to support runtime trust and suggesting a roadmap for their integration into future O-RAN specifications.

The main contributions of this paper are as follows:
\begin{enumerate}
    \item rApp/xApp attestation is introduced as a new O-RAN security use case to address the lack of runtime integrity verification in current specifications.

    \item A standards-aligned attestation workflow is designed and mapped to relevant O-RAN working groups and interfaces, providing a concrete integration and standardization path.

    \item The feasibility of the proposed approach is demonstrated through a lightweight prototype integrated into a Near-RT RIC platform, showing that runtime attestation can be performed with millisecond-scale overhead.

    \item Key challenges and open issues in deploying runtime attestation in O-RAN are systematically identified, together with directions for scalable and standardized integration.
\end{enumerate}

The remainder of this paper is organized as follows. Section~\ref{background} reviews the relevant O-RAN security specifications and identifies the runtime integrity gap. Section~\ref{sec:usecase} presents the proposed rApp/xApp attestation use case. Section~\ref{standards} discusses its alignment with O-RAN working groups and related telecom security frameworks. Section~\ref{results} describes the prototype implementation and evaluation results. Section~\ref{openissues} outlines remaining challenges and future directions, and Section~\ref{conclusion} concludes the paper.

\section{Standards Context: Background and Gaps}\label{background}

The O-RAN Alliance has defined a comprehensive set of security specifications under WG11 covering O-Cloud security, application onboarding, authentication, lifecycle management, and interface protection for rApps/xApps and RIC components. These specifications collectively establish baseline security controls, including application registration, identity management, secure communication, and conformance testing. The O-Cloud security assessment further recommends remote attestation for establishing trust in O-Cloud components and service deployments, including VM/container measurement at launch and while in use. In addition, several study items explicitly recognize the risk of compromised rApps/xApps and highlight runtime integrity as a potential concern.

Table~\ref{tab:oran-specs} summarizes the scope, coverage, and limitations of the most relevant specifications, highlighting the absence of concrete mechanisms for runtime integrity verification.

These specifications demonstrate that runtime integrity is a recognized concern in O-RAN security, yet no concrete RIC-native mechanism or implementation guidance exists for runtime rApp/xApp attestation. This gap motivates rApp/xApp attestation as a practical, standardizable approach to runtime security for the RIC.

\begin{table*}[h]
\centering
\caption{\small Summary of relevant O-RAN security specifications and identified gaps.}
\label{tab:oran-specs}
\begin{tabularx}{\textwidth}{|p{2.4cm}|p{2.4cm}|p{7cm}|X|}
\hline
\textbf{Specification} & \textbf{Focus Area} & \textbf{Coverage} & \textbf{Gap} \\
\hline
WG11 SRCS TS \cite{oran-srcs}
&
Security requirements (baseline controls)
&
Registration, identity management, API authorization
&
No requirement that mandates rApp/xApp runtime integrity verification
\\
\hline

WG11 AppLCM Security TR \cite{oran-LCM}
&
Application lifecycle security (operation phase)
&
Identifies runtime threats; proposes cryptographic hash verification as a potential requirement; discusses feasibility limits
&
Acknowledged issue, but no informative mechanism due to feasibility concerns
\\
\hline

WG11 Threat Modeling TR \cite{oran-Threat}
&
Comprehensive threat catalog
&
Identifies threats across O-RAN; compromised rApps/xApps highlighted as a major risk
&
No defined countermeasures or runtime defenses
\\
\hline

WG11 Security Test Specs TS \cite{oran-Sec-Test}
&
Security validation
&
Defines onboarding, package signing and verification, OAuth 2.0 client authorization, and lifecycle tests (e.g., signature validation and secure decommissioning) for rApps/xApps
&
No runtime integrity verification or attestation testing
\\
\hline

WG11 ZTA TR \cite{zta}
&
Zero Trust security for O-RAN
&
Identifies security gaps and additional security controls across xApp/Near-RT RIC and rApp/Non-RT RIC ecosystems
&
No runtime attestation mechanism for verifying rApp/xApp integrity during operation
\\
\hline

WG11 SMO Security Analysis TR \cite{oran-SMO-sec}
&
SMO, Non-RT RIC, and rApp security
&
Identifies rApp management and software onboarding as security-relevant; analyzes threats from compromised or misbehaving rApps
&
No runtime integrity verification or attestation mechanism for rApps/xApps
\\
\hline

WG11 Near-RT RIC \& xApps TR \cite{oran-Near-xApps}
&
Near-RT RIC and xApps threats
&
Identifies compromised xApps as a key security concern and recommends onboarding safeguards
&
No runtime integrity verification or attestation mechanism for xApps after deployment
\\
\hline
WG11 O-Cloud Security TR \cite{oran-OCloud-sec}
&
O-Cloud security and threat assessment
&
Recommends remote attestation of O-Cloud components,
including VM/container measurement at launch and while in use
&
No RIC-native runtime attestation procedure for rApps/xApps
\\
\hline
\end{tabularx}
\end{table*}

\section{Proposed Use Case: rApp/xApp Attestation}
\label{sec:usecase}

To enable runtime assurance of rApps and xApps in O-RAN, we propose \textit{rApp/xApp attestation} as a runtime integrity verification use case within the RIC. The key idea is that the RIC continuously or periodically verifies the integrity of deployed applications by requesting and validating runtime evidence against trusted reference measurements. This enables operators to maintain assurance over application behavior throughout their lifecycle, complementing existing onboarding and access control mechanisms.

\begin{figure*}[t]
    \centering
    \includegraphics[width=\textwidth]{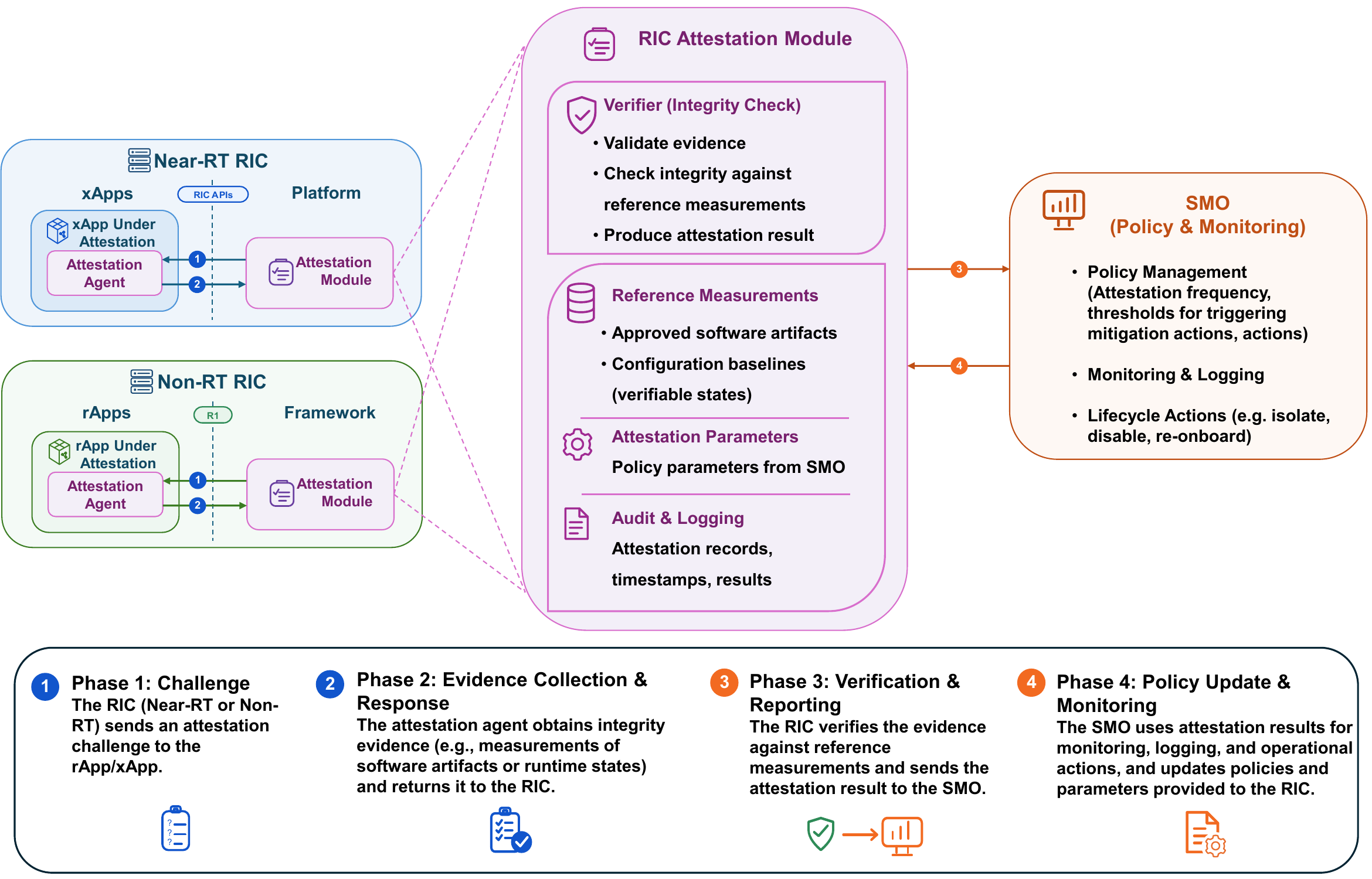}
    \caption{\small 
    Architecture of the proposed rApp/xApp attestation use case in O-RAN. 
    The RIC (Near-RT and Non-RT) performs runtime integrity verification of applications through an attestation function, interacting with rApps/xApps via standard interfaces and reporting outcomes to the SMO for policy-driven monitoring and control.
    }
    \label{fig:architecture}
\end{figure*}

Attestation interactions cross the RIC application interfaces, using RIC APIs in the Near-RT RIC and the R1 interface in the Non-RT RIC. Fig.~\ref{fig:architecture} illustrates the architecture and operational phases of the proposed attestation use case. At a high level, the attestation process follows a closed-loop interaction between the RIC, the application (rApp/xApp), and the SMO. The RIC initiates an attestation challenge, to which the application responds by obtaining integrity evidence derived from its runtime state. This evidence is then verified within the RIC against trusted reference measurements. Based on the verification outcome, the RIC reports the result to the SMO, which provides policy-driven guidance for monitoring and mitigation actions. Although attestation policies are provisioned by the SMO, the operational cycle is executed within the RIC, enabling local and timely verification.

The main functional roles in this architecture can be summarized as follows:
\begin{itemize}
    \item \textit{RIC (Near-RT and Non-RT):} acts as the verifier, realized through an internal \textit{attestation module} that issues attestation challenges, validates received evidence against reference measurements, maintains attestation records, and enforces mitigation actions when required.
    
    \item \textit{rApps/xApps:} act as attestation targets, equipped with a lightweight \textit{attestation agent} that responds to challenges by obtaining and returning integrity evidence derived from their software artifacts or runtime state.
    
    \item \textit{SMO:} provides attestation policies (e.g., frequency and operational thresholds for triggering mitigation actions, such as repeated verification failures or excessive response delays) and consumes attestation results for monitoring, logging, and operational control.
\end{itemize}

\subsection{Attestation Model and Data Abstraction}

The proposed use case follows an integrity-based attestation model, where measurements of software and configuration states are validated against trusted reference values. Such references should correspond to an approved application version and configuration and be updated through authorized lifecycle changes, ensuring that runtime evidence is evaluated against the appropriate known-good state. This approach is lightweight, directly verifiable, and suitable for periodic runtime checking in RIC environments. The attestation process relies on three types of data: (i) reference measurements, representing trusted software and configuration baselines maintained within the RIC; (ii) runtime evidence, obtained from rApps/xApps in response to attestation challenges; and (iii) metadata and logs, including timestamps and verification outcomes used for auditing and policy enforcement. 

While integrity-based verification provides a practical foundation, complementary mechanisms such as behavioral evidence (e.g., system calls, execution traces, or control actions) may further enhance detection coverage. However, such approaches are more context-dependent and are therefore considered as extensions rather than core components of the proposed use case. It is discussed in more detail in Section~\ref{openissues}

\subsection{Threat Model}

The attacker is assumed to compromise the software environment of a deployed rApp or xApp after successful onboarding and registration. Possible attack vectors include binary manipulation, code injection, configuration tampering, or unauthorized updates. The attacker’s goal is to alter control logic or data flows while remaining undetected by existing onboarding and interface protections.

Consistent with existing O-RAN security analyses, the Non-RT framework and Near-RT RIC platform and their attestation module are assumed to remain trusted, with securely maintained reference measurements. The threat model also considers attempts to forge attestation responses or replay outdated evidence, requiring the attestation mechanism to detect both static and runtime modifications in a timely manner.

\subsection{Mitigation Actions}
\label{mitigation}

Upon detecting integrity violations, the RIC applies mitigation actions that balance network protection with service continuity. Rather than relying on a single response, mitigation can be hierarchical and policy-driven:
\begin{itemize}
    \item \textit{Isolation:} temporarily remove the affected rApp/xApp from control loops while keeping it under observation.
    \item \textit{Privilege reduction:} restrict access to APIs or configuration capabilities until integrity is re-established.
    \item \textit{Quarantine and validation:} move the application to a controlled state for further inspection or re-attestation.
    \item \textit{Controlled deactivation:} disable the application via lifecycle management when necessary and notify the SMO.
    \item \textit{Recovery and re-onboarding:} restore the application after validation through standard onboarding procedures.
\end{itemize}

These actions are governed by SMO-configured policies, which define how attestation outcomes are translated into mitigation decisions. Such policies may consider factors such as repeated verification failures or the severity of detected violations. Standardizing policy-driven mitigation procedures would further enhance interoperability and consistent security enforcement across multi-vendor O-RAN deployments.

\section{Standards Alignment and Roadmap}
\label{standards}

The proposed rApp/xApp attestation use case complements and extends existing O-RAN specifications by addressing the absence of runtime integrity verification identified in Section~\ref{background}. It builds on current architectural components and interfaces, enabling integration without introducing new entities. In the absence of standardization, such mechanisms are likely to be implemented in a vendor-specific manner, leading to fragmentation and reduced interoperability.

\subsection{Architectural Integration within the RIC (WG1, WG2, WG3)}
\label{ArchInteg}

Attestation can be realized within the Non-RT RIC framework and the Near-RT RIC platform, which host rApps and xApps, respectively, and support service-based interaction models. This aligns with WG1, where use cases are defined and integrated into the overall O-RAN system design. Extending the \textit{O-RAN Use Cases Detailed Specification} \cite{oran-Use-Cases} to include runtime attestation represents a natural evolution of existing security-related scenarios.

At the interface level, the proposed approach can be realized through dedicated services, allowing both the RIC and rApps/xApps to act as either service providers or consumers. In the Non-RT RIC (WG2), attestation can be supported over the R1 interface by introducing a dedicated service in which the rApp provides integrity evidence while the RIC issues challenges and verifies responses. In the Near-RT RIC (WG3), although communication between xApps and the Near-RT RIC platform may differ across RIC implementations, the O-RAN specifications define a Near-RT RIC API framework supporting service-based interactions between xApps and the Near-RT RIC platform, including API registration and discovery \cite{oran-RICAPI}. Accordingly, a dedicated xApp attestation service and its corresponding API can be defined within the WG3 Near-RT RIC API framework, enabling challenge--response exchange and reporting between the RIC and xApps. This provides a common
standardization path while allowing implementation-specific communication mechanisms to differ across Near-RT RIC implementations.

In both contexts, rApps/xApps act as service providers of attestation evidence, while the RIC acts as the service consumer responsible for initiating attestation and validating responses. As a result, attestation can be incorporated without introducing new functional entities, requiring only the definition of the corresponding attestation services and APIs.

\subsection{Lifecycle and Operational Integration (WG10 and SMO Coordination)}

WG10 specifies the \textit{O-RAN Operations and Maintenance Architecture} \cite{oran-OAM-Arch}, including SMO-driven lifecycle management via the O1 interface. Within this framework, attestation results can be reported from the RIC to the SMO, enabling their use in monitoring, fault management, and policy-driven control.

SMO policies can govern key operational parameters of attestation, such as frequency, triggering conditions, thresholds for failure, and corresponding mitigation actions. By integrating attestation outcomes into existing lifecycle workflows, operators can enforce consistent and automated responses to runtime integrity violations without introducing new management frameworks.

\subsection{Security Integration (WG11)}

From a security perspective, WG11 specifications can be extended to include requirements for runtime integrity verification, including the provisioning of trusted reference measurements (e.g., approved software and configuration baselines) during onboarding. These references enable subsequent verification of runtime evidence obtained from rApps/xApps.

In addition, WG11 can define supporting procedures and validation mechanisms for attestation, such as challenge–response verification, freshness guarantees, and reporting of outcomes, enabling consistent implementation and evaluation across vendors. Attestation extends the WG11 security framework with runtime integrity verification, complementing and operating alongside existing mechanisms for identity management, authentication, and certificate handling.

Overall, this extends the O-RAN security framework from pre-deployment assurance to continuous runtime trust validation, while remaining aligned with existing architectural and operational models.

Accordingly, standardization should focus on defining the attestation services and APIs, evidence and reference requirements, and validation procedures, while allowing different attestation mechanisms to realize these requirements. The remaining technical challenges for practical adoption are discussed in Section~\ref{openissues}.

\subsection{Relation to Existing Telecom Security Frameworks}

Existing telecom security frameworks provide useful context but do not
directly address runtime integrity of RIC applications. ETSI
\textit{NFV-SOL 004}, which specifies VNF package and archive requirements,
includes integrity protection through package signing and verification.
Similarly, 3GPP \textit{Security Assurance Specifications (SCAS)}, such as
\textit{TS 33.117}, and the GSMA \textit{Network Equipment Security Assurance
Scheme (NESAS)} focus primarily on security assurance and conformance testing.
These approaches provide relevant foundations but require adaptation to the
service-based, control-loop-driven, and heterogeneous COTS environment of O-RAN.

\section{Prototype and Results} \label{results}

The proposed attestation use case applies to both rApps in the Non-RT RIC and xApps in the Near-RT RIC. However, these platforms operate under significantly different timing constraints. For this reason, we prioritize feasibility evaluation where requirements are most stringent. Our hypothesis is that if runtime attestation introduces sufficiently low overhead to remain practical within Near-RT RIC operations, then the same approach will be inherently feasible for rApps in the Non-RT RIC, which is less time-critical.

We therefore implement and evaluate the attestation mechanism only for xApps, while noting that the design and workflow generalize directly to rApps as described in Section~\ref{sec:usecase}. This section presents both the prototype realization of the proposed attestation mechanism and its performance evaluation, providing insight into its practical deployment in the O-RAN architecture. The prototype implements a lightweight hash-based attestation mechanism, where integrity is verified through cryptographic digests of application binaries combined with challenge–response interaction.

The prototype builds upon the xApp integrity check module from our prior work \cite{Towards}. However, that implementation did not optimize the verification pipeline and introduced avoidable overhead. Runtime attestation requires repeated access to application binaries and reference measurements, which introduces non-negligible I/O and computation overhead. In RIC environments, excessive overhead may limit the feasibility of frequent verification. Therefore, optimizing the verification process is essential to ensure that attestation can be performed efficiently without interfering with normal operation.

The most impactful improvement was redesigning the memory access routine and implementing a more efficient file-reading mechanism, reducing verification overhead by $\sim$30\%. This was achieved by eliminating repeated buffered reads and mapping the reference file directly into memory, reducing copy overhead and enabling more efficient access during hash computation. This optimization enables faster loading and hashing of the reference image, particularly for larger digests. We further extended the module to evaluate multiple cryptographic hash functions, allowing us to quantify execution cost and study how the optimized implementation scales with digest size.

\subsection{Prototype and Evaluation Setup}

To validate feasibility under Near-RT constraints, we implemented a proof-of-concept xApp attestation prototype on a FlexRIC-based Near-RT RIC platform, reflecting a realistic deployment where xApps operate under latency-sensitive conditions. The prototype implements a lightweight hash-based attestation mechanism, where integrity is verified through cryptographic digests of application binaries, computed over their memory-resident executable content, combined with a challenge–response interaction.

\begin{figure*}[t]
    \centering
    \includegraphics[width=0.85\linewidth]{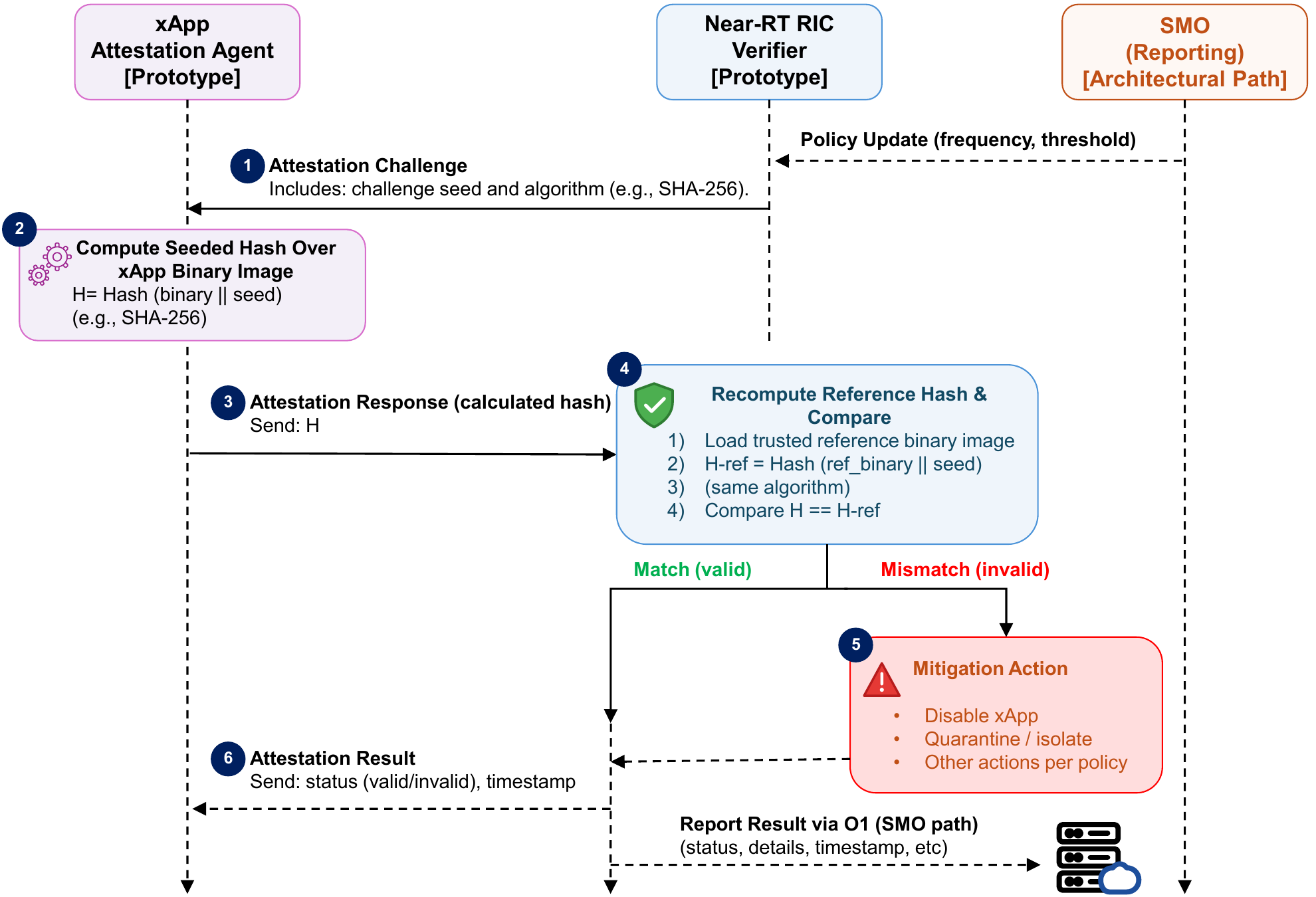}
    \caption{\small Sequence of the hash-based runtime attestation procedure in the implemented prototype.}
    \label{fig:attestation-sequence}
\end{figure*}

The interaction workflow is illustrated in Fig.~\ref{fig:attestation-sequence}. A custom xApp was extended with an attestation agent that responds to challenges issued by the attestation module in the Near-RT RIC, which acts as the verifier and maintains a trusted baseline image of the xApp. Upon connection, the xApp provides its metadata and receives an attestation challenge. The attestation agent computes a hash over its executable content combined with a challenge seed and returns the result as integrity evidence. The verifier recomputes the expected hash over the trusted reference using the same seed and compares the results to determine integrity. This seed-based challenge–response protocol ensures freshness and protects against replay attacks while remaining lightweight and suitable for periodic runtime verification.

In the prototype, the attestation functionality is exposed through a dedicated API implemented for this work. For standardization, a corresponding xApp attestation service and API could be defined within the WG3 Near-RT RIC API framework, as discussed in Section~\ref{ArchInteg}.

The prototype consumes a set of well-defined input data and produces corresponding outputs, summarized in Table~\ref{tab:xapp-io}. This mapping also illustrates how attestation interactions integrate with existing O-RAN interfaces.

We evaluated the prototype using an xApp with a binary footprint of approximately 10~MB, representative of lightweight Near-RT RIC applications. Each attestation cycle consists of challenge issuance, hash computation at the xApp, response transmission, and verification at the RIC. By structuring the evaluation around this isolated attestation cycle, the measured latency directly reflects the overhead introduced by the attestation mechanism, without being conflated with other RAN processing tasks.

The xApp computes hashes over its readable-executable (\texttt{r-xp}) memory mappings, while the RIC recomputes the digest over a stored reference image. Writable regions are excluded because their contents may
legitimately change at runtime and therefore do not provide a stable
reference for direct hash comparison. 

Experiments were executed on an Intel\textregistered{} Core\texttrademark{} i7-14700 CPU @ 2.10~GHz, with 32~GB RAM, running Ubuntu 24.04 (64-bit).

\begin{table*}[t]
\centering
\caption{\small Required Input and Output Data for xApp Attestation.}
\label{tab:xapp-io}
\renewcommand{\tabularxcolumn}[1]{>{\raggedright\arraybackslash}m{#1}}

\begin{tabularx}{\textwidth}{|m{1cm}|m{3.2cm}|m{2.6cm}|m{2.0cm}|X|}
\hline
\textbf{Type} & \textbf{Interface} & \textbf{Source} & \textbf{Target} & \textbf{Description} \\
\hline
\multirow{2}{*}{\textbf{Input}} 
& O1 & SMO & Near-RT RIC platform & Attestation parameters such as frequency, thresholds, and mitigation policies. \\
\cline{2-5}
& Near-RT RIC API – Future xApp attestation API & Near-RT RIC platform & xApp & Challenge seed ensuring freshness and preventing replay attacks. \\
\hline
\multirow{2}{*}{\textbf{Output}} 
& O1 & Near-RT RIC platform & SMO & Validation results, timestamps, alerts, and escalation events. \\
\cline{2-5}
& Near-RT RIC API – Future xApp attestation API & xApp & Near-RT RIC platform & Attestation response including calculated hash. \\
\hline
\end{tabularx}
\end{table*}

\subsection{Results}

\begin{figure}[h!]
    \centering
    \includegraphics[width=\linewidth]{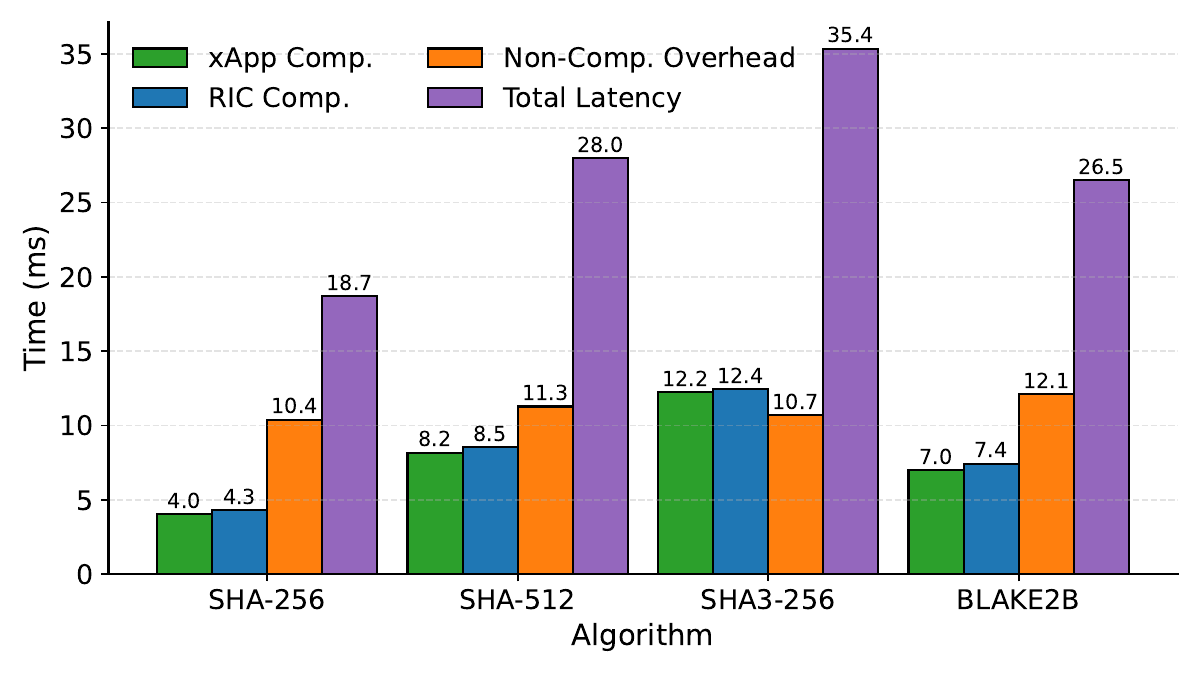}
    \caption{\small xApp attestation latency.}
    \label{fig:attest-latency}
\end{figure}

We benchmarked attestation using SHA-256, SHA-512, SHA3-256, and BLAKE2b (512-bit), comparing classic SHA-2 hashes with more manipulation-resistant designs. Each algorithm ran across 50 rounds. The first round exhibited only a modest cold-start increase of approximately 3--4 ms compared with the subsequent rounds and was excluded from the reported steady-state averages. Fig.~\ref{fig:attest-latency} reports the latency breakdown of the attestation cycle, including xApp-side hash computation, RIC-side verification, and additional overhead, along with the total latency. The total latency corresponds to the sum of these components. The overhead component captures the non-computational costs of attestation, including communication between the xApp and RIC, protocol handling, and scheduling effects. In this decomposition, the dominant contributors to latency are the hashing and verification steps, while the overhead remains relatively stable across algorithms. These measured components collectively represent the additional processing cost introduced by attestation relative to a baseline system without attestation.

Overhead remains nearly constant across algorithms, with small growth tied to digest size. SHA-512 is roughly twice as slow as SHA-256, and SHA3-256 is the slowest despite identical output length. SHA3-256 and BLAKE2b-512 provide stronger resistance against length-extension and crafted-collision attacks, but at a higher computation cost.

Practically, SHA-256 remains the most suitable algorithm for frequent attestation, while SHA-512 increases digest size and collision resistance with moderate added cost. SHA3-256 and BLAKE2b-512 offer stronger robustness at higher cost, with BLAKE2b providing a favorable balance between performance and security. All algorithms complete within millisecond-scale timing, confirming feasibility for periodic runtime attestation in Near-RT RIC.

Using the same prototype and attestation procedure, we further evaluated
its detection capability against selected runtime attack scenarios.
As shown in Table~\ref{tab:attack}, these cases illustrate the detection
boundary of the prototype: modifications affecting the measured executable
content are detected, whereas changes outside the measured state remain
beyond the coverage of hash-based attestation.

\begin{table}[t]
\centering
\caption{\small {Runtime attack coverage.}}
\label{tab:attack}
\begin{tabular}{|p{5.5cm}|c|}
\hline
\textbf{Test case} & \textbf{Result} \\
\hline
Existing executable-code modification & Detected \\
\hline
Code injection into new \texttt{r-xp} mapping & Detected \\
\hline
Modification outside measured \texttt{r-xp} state & Not detected \\
\hline
\end{tabular}
\end{table}

Although attestation operates outside the main control loop of Near-RT RIC and is not subject to strict real-time constraints, it is executed periodically and introduces additional computational load on the hosting platform. Excessive delay could indirectly impact the control loop execution if verification interferes with ongoing xApp processing. However, the measured latencies remain in the millisecond range, which is small relative to typical RIC processing timescales, indicating that such effects are limited in practice. This impact can be further mitigated through appropriate scheduling and attestation frequency control, but their effectiveness under representative RAN workloads requires further evaluation.

\subsection{Implications}

The evaluation demonstrates that integrating runtime attestation within the RIC introduces only millisecond-scale overhead, confirming its feasibility under Near-RT constraints. The prototype shows that runtime xApp attestation is technically feasible under the evaluated Near-RT RIC conditions. By introducing only limited additional latency in the evaluated setup, the mechanism supports periodic integrity verification alongside time-sensitive control operations when appropriately scheduled. 

This positions attestation as a practical extension to existing O-RAN security mechanisms, while leaving room for future enhancements such as behavioral monitoring or hardware-assisted verification.

\section{Challenges and Future Directions} \label{openissues}

While the prototype demonstrates the feasibility of runtime xApp attestation, several challenges remain before such mechanisms can be deployed at scale in O-RAN systems. These include concerns already highlighted in Section 6.9 of the O-RAN.WG11.TR.AppLCM-Security report \cite{oran-LCM}, as well as additional issues identified in this work. Together, they outline both the limitations of current approaches and the directions required for practical deployment and standardization.

\textit{Security of the monitoring entity:}
A monitoring entity verifying application integrity must itself remain protected, as compromise would invalidate attestation results. In the proposed mechanism, the verifier is part of the trusted Non-RT framework and Near-RT RIC platform and inherits its protection mechanisms. While this aligns with existing O-RAN assumptions, stronger trust anchors (e.g., TPMs or TEEs) could further enhance protection where available, although their deployment may be constrained in virtualized and COTS environments.

\textit{Defining known-good runtime states:}
A key challenge lies in defining what constitutes an ``untampered'' application during runtime. In this context, a known-good state is best interpreted as a verifiable software integrity state matching trusted reference artifacts (e.g., binaries or configuration baselines), rather than application behavior or network outcomes, which are context-dependent. However, runtime memory and configuration values may evolve legitimately, particularly in AI/ML-driven xApps, making stable reference selection non-trivial. Static hashing alone may therefore be insufficient, motivating richer verification approaches that combine integrity validation with complementary runtime observations.

\textit{Stealthy attacks and limitations of software-based attestation:}
Attackers may evade detection by leaving the measured executable state unchanged while executing malicious logic outside that state. While seed-based attestation prevents replay attacks, it cannot fully address such stealthy behavior. Hash-based approaches are lightweight and cloud-friendly but rely on weak roots of trust. Complementary mechanisms, such as timing-based heuristics or behavioral monitoring (e.g., system calls, execution traces, or control anomalies), can improve detection coverage. Hash comparison itself is deterministic; however, legitimate changes to measured state may cause mismatches if trusted references are outdated. Reference lifecycle management is therefore necessary to distinguish authorized changes from integrity violations.

\textit{Hybrid attestation mechanisms:}
To address these limitations, hybrid approaches combining software-based attestation with hardware-assisted trust anchors (e.g., TPMs or TEEs) offer a promising direction. Hardware-assisted methods may be applied selectively to critical functions, while lightweight software attestation scales to broader deployments. Such designs improve trust guarantees but introduce challenges in deployment, interoperability, standardizability, and system complexity.

\textit{Performance and scalability:}
Aggregate attestation load grows with the number of applications,
verification frequency, and per-attestation cost. Concurrent checks may
therefore cause resource contention and indirectly affect Near-RT RIC
processing. Staggered scheduling, application prioritization, and adaptive
frequency can limit such overhead. Multi-xApp scalability and performance
under heavy RAN workloads remain to be experimentally evaluated.

\textit{Mitigation actions and lifecycle integration:}
While mitigation strategies (e.g., isolation, privilege reduction, and controlled deactivation) can be triggered upon attestation failure, their conditions and integration into lifecycle management workflows are not yet fully defined. Standardizing these mechanisms, including their coordination with SMO-driven policies, is essential to ensure consistent behavior across vendors while preserving service continuity.

\textit{Expanding the scope across the O-RAN ecosystem:}
Although this work focuses on rApps/xApps, similar integrity challenges exist for VNFs and cloud-native workloads. Extending attestation across these layers could enable a unified trust framework for disaggregated RAN systems, forming the basis of a broader ``trust fabric'' aligned with O-RAN’s open and programmable architecture.

Overall, these challenges and directions indicate that while runtime attestation is technically feasible, its large-scale adoption requires advances in hybrid trust models, complementary monitoring, and adaptive orchestration. Addressing these aspects will be key to transforming rApp/xApp attestation from a prototype capability into a practical and standardized feature of O-RAN systems.

\section{Conclusion} \label{conclusion}

O-RAN programmability enables flexible and intelligent RAN control, but it also creates a need to ensure that deployed rApps and xApps remain trustworthy during operation. This paper introduced rApp/xApp attestation as a RIC-native security use case for runtime integrity verification. The proposed use case complements existing onboarding, authentication, and interface protection mechanisms by extending trust assurance into the operational phase.

We mapped the attestation workflow to relevant O-RAN Alliance working groups and interfaces, showing how attestation can be realized through controlled extensions to R1, RIC APIs, and SMO-based lifecycle coordination. A hash-based prototype on a Near-RT RIC platform demonstrated that runtime attestation can be performed with millisecond-scale overhead, supporting its feasibility for periodic verification in latency-sensitive environments. The remaining challenges include securing the verifier, defining stable known-good states, scaling attestation across many applications, integrating mitigation actions into lifecycle workflows, and exploring hybrid mechanisms that combine software-based checks with hardware-assisted trust where available. Addressing these issues can help transform rApp/xApp attestation from a prototype capability into a standardized security mechanism for secure and interoperable O-RAN deployments, providing a foundation for future O-RAN Alliance standardization efforts in runtime trust assurance.

\section*{Acknowledgment}
This work was supported by UKRI (UK ID: 640240), CDTI (ES), ANI (PT ID: COMPETE2030-FEDER-01602800), and TUBITAK (TR ID: 9240013) within the EUREKA CELTIC-NEXT project 6G-SMART, C2023/2-20.


\bibliographystyle{IEEEtran}
\bibliography{MyRef.bib}

@techreport{oran-srcs,
  title        = {{O-RAN Security Requirements and Controls Specification}},
  institution  = {O-RAN Alliance},
  type         = {Technical Specification},
  number       = {O-RAN.WG11.TS.SRCS.0-R005-v15.00},
  year         = {2026},                 
  note         = {Available at: \url{https://specifications.o-ran.org/specifications}}
}

@techreport{oran-LCM,
  title        = {{Study on Security for Application Lifecycle Management}},
  institution  = {O-RAN Alliance},
  type         = {Technical Report},
  number       = {O-RAN.WG11.TR.AppLCM-Security-R004-v04.00},
  year         = {2025},                 
  note         = {Available at: \url{https://specifications.o-ran.org/specifications}}
}

@techreport{oran-Threat,
  title        = {{O-RAN Security Threat Modeling and Risk Assessment}},
  institution  = {O-RAN Alliance},
  type         = {Technical Report},
  number       = {O-RAN.WG11.TR.Threat-Modeling-R005-v09.00},
  year         = {2026},                 
  note         = {Available at: \url{https://specifications.o-ran.org/specifications}}
}

@techreport{ZTA,
  title        = {{Study on Zero Trust Architecture for O-RAN}},
  institution  = {O-RAN Alliance},
  type         = {Technical Report},
  number       = {O-RAN.WG11.TR.ZTA-R005-v06.00},
  year         = {2026},                 
  note         = {Available at: \url{https://specifications.o-ran.org/specifications}}
}

@techreport{oran-Near-xApps,
  title        = {{Study on Security for Near Real Time RIC and xApps}},
  institution  = {O-RAN Alliance},
  type         = {Technical Report},
  number       = {O-RAN.WG11.Security-Near-RT-RIC-xApps-TR.0-R005-v07.00},
  year         = {2026},                 
  note         = {Available at: \url{https://specifications.o-ran.org/specifications}}
}

@techreport{oran-SMO-sec,
  title        = {{Study on Security for 
Service Management and Orchestration 
(SMO)}},
  institution  = {O-RAN Alliance},
  type         = {Technical Report},
  number       = {O-RAN.WG11.TR.SMO-Security-Analysis.0-R005-v08.00},
  year         = {2026},                 
  note         = {Available at: \url{https://specifications.o-ran.org/specifications}}
}

@techreport{oran-Sec-Test,
  title        = {{O-RAN Security Test Specifications}},
  institution  = {O-RAN Alliance},
  type         = {Technical Specification},
  number       = {O-RAN.WG11.TS.STS-R005-v13.00},
  year         = {2026},                 
  note         = {Available at: \url{https://specifications.o-ran.org/specifications}}
}

@techreport{oran-Use-Cases,
  title        = {{Use Cases Detailed Specification}},
  institution  = {O-RAN Alliance},
  type         = {Technical Specification},
  number       = {O-RAN.WG1.TS.Use-Cases-Detailed-Specification-R005-v20.00},
  year         = {2026},                 
  note         = {Available at: \url{https://specifications.o-ran.org/specifications}}
}

@techreport{oran-OAM-Arch,
  title        = {{O-RAN Operations and Maintenance Architecture}},
  institution  = {O-RAN Alliance},
  type         = {Technical Specification},
  number       = {O-RAN.WG10.TS.OAM-Architecture-R005-v18.00},
  year         = {2026},                 
  note         = {Available at: \url{https://specifications.o-ran.org/specifications}}
}

@ARTICLE{ztran,
  author={Abdalla, Aly S. and Moore, Joshua and Adhikari, Nisha and Marojevic, Vuk},
  journal={IEEE Wireless Communications}, 
  title={{ZTRAN: Prototyping zero trust security xApps for open radio access network deployments}}, 
  year={2024},
  volume={31},
  number={2},
  pages={66-73},
  doi={10.1109/MWC.001.2300419}
}

@ARTICLE{Towards,
  author={Alimohammadi, Hamed and Mayhoub, Samara and Chatzimiltis, Sotiris and Shojafar, Mohammad and Bhutta, Muhammad Nasir Mumtaz},
  journal={IEEE Open Journal of the Communications Society}, 
  title={{Toward a Multi-Layer Defence Framework for Securing Near-Real-Time Operations in Open RAN}}, 
  year={2026},
  volume={7},
  number={},
  pages={480-497},
  doi={10.1109/OJCOMS.2025.3650736}}

@ARTICLE{OJCOMS,
  author={Hung, Cheng-Feng and Chen, You-Run and Tseng, Chi-Heng and Cheng, Shin-Ming},
  journal={IEEE Open Journal of the Communications Society}, 
  title={{Security Threats to xApps Access Control and E2 Interface in O-RAN}}, 
  year={2024},
  volume={5},
  number={},
  pages={1197-1203},
  doi={10.1109/OJCOMS.2024.3364840}}

@ARTICLE{TDSC,
  author={Atalay, Tolga O. and Maitra, Sudip and Stojadinovic, Dragoslav and Stavrou, Angelos and Wang, Haining},
  journal={IEEE Transactions on Dependable and Secure Computing}, 
  title={{An OpenRAN Security Framework for Scalable Authentication, Authorization, and Discovery of xApps With Isolated Critical Services}}, 
  year={2025},
  volume={22},
  number={3},
  pages={2873-2890},
  doi={10.1109/TDSC.2024.3522218}}

@techreport{oran-ocloud-sec,
  title        = {{Study on Security for O-Cloud}},
  institution  = {O-RAN Alliance},
  type         = {Technical Report},
  number       = {O-RAN.WG11.TR.O-CLOUD-Security.0-R005-v09.00},
  year         = {2026},                 
  note         = {Available at: \url{https://specifications.o-ran.org/specifications}}
}

@techreport{oran-RICAPI,
  title        = {{Near-RT RIC APIs}},
  institution  = {O-RAN Alliance},
  type         = {Technical Specification},
  number       = {O-RAN.WG3.TS.RICAPI-R005-v03.00},
  year         = {2026},                 
  note         = {Available at: \url{https://specifications.o-ran.org/specifications}}
}


\section*{Biographies}
\vspace{-10mm}
\begin{IEEEbiographynophoto}{Hamed Alimohammadi}
 is a senior research fellow in mobile networks at the University of Surrey’s 6G Innovation Centre, specializing in Open RAN self-organization and security. He previously spent nearly a decade in mobile network governance and engineering roles. He holds a PhD in Computer Engineering from Razi University, Iran. At Surrey, he has contributed to flagship research programs including 6G-SMART and HiPer-RAN, where he was part of the team recognized with the UK Government’s Future Network “Incremental Innovative” Award in 2025. His research interests span Open RAN security, machine learning for networks, and high-performance computing.
\end{IEEEbiographynophoto}
\vspace{-30pt}
\begin{IEEEbiographynophoto}{Burcu Şahin}
 is an O-RAN architect with over 15 years of experience in research, software development, product architecture, and standardization across wireless networks. She holds BSc and MSc degrees in Computer Science from Bilkent University and a PhD in Cognitive Science from Middle East Technical University. Burcu has contributed to R\&D, product and standardization roles at Turk Telekom, Argela, and Juniper Networks, and currently works at Nokia focused on RAN Intelligent Controller product management and architecture. An active O-RAN Alliance standards contributor since 2018, she serves as rapporteur of E2SM-CCC and authored more than 20 patents.
\end{IEEEbiographynophoto}
\vspace{-30pt}
\begin{IEEEbiographynophoto}{Arda Akman}
has over 30 years of R\&D expertise in wireless networks. Currently, as R\&D Director at Nokia, he is leading the product management and architecture of RAN Intelligent Controller (RIC). His previous roles include Senior Director of Engineering at Juniper Networks, Argela/Netsia, Turk Telekom R\&D, Ixia and Nortel Networks. A very active contributor in O-RAN Alliance since 2018, he is leading the Use Case Task Group, Network Slicing Task Group and is also the rapporteur for 5 different specifications. He holds BSc and MSc degrees in Electronics Engineering and is the author of over 30 patent applications.
\end{IEEEbiographynophoto}
\vspace{-30pt}
\begin{IEEEbiographynophoto}{Chuan Heng Foh}
(Senior Member, IEEE) received his M.Sc. from Monash University, Australia, in 1999, and his Ph.D. from The University of Melbourne in 2002. He briefly lectured at Monash before joining Nanyang Technological University, Singapore, as an Assistant Professor in 2002. He joined the University of Surrey in 2012 and is currently an Associate Professor there. He has authored over 200 refereed publications in international journals and conferences. His research interests include protocol design, machine learning applications, IoT, vehicular, wireless LANs, mobile ad hoc networks, 5G/6G systems, and Open RAN. He has held several IEEE leadership and editorial roles.
\end{IEEEbiographynophoto}

\vspace{-30pt}
\begin{IEEEbiographynophoto}{Periklis Chatzimisios}
(Senior Member, IEEE) is currently a Professor at the International Hellenic University, Greece, and a Research Professor at the University of New Mexico, USA. His research interests include next-generation wireless communications, standardization, the Internet of Things (IoT), legal and ethical issues of artificial intelligence (AI), security, and robotics. He is the Vice-Chair of Working Group 1 and a Board Member of the one6G Association. He is currently an IEEE Communications Society Distinguished Lecturer. He has been included in Stanford University's list of the world's most influential scientists from 2020 to 2025. He serves as an Associate Editor-in-Chief of IEEE Communications Standards Magazine.
\end{IEEEbiographynophoto}
\vspace{-30pt}
\begin{IEEEbiographynophoto}{Mohammad Shojafar}
is an Associate Professor in Network Security, an Intel Innovator, a Senior IEEE member, an ACM Distinguished Speaker, and a Higher Education Academy Fellow at the 6GIC, University of Surrey. He has secured around £3M as Principal Investigator across EU and UK projects. With the Surrey team, he received the “Incremental Innovative” Future Network Award for the DSIT/UKTIN HiPer-RAN project in 2025 and a Best Paper Award at the IEEE CSNet Conference in 2023. He is also the author of three Springer books and serves as Associate Editor for IEEE TNSM, IEEE TITS, IEEE TGCN, and IEEE TCE.
\end{IEEEbiographynophoto}

\vfill
\immediate\write18{texcount -inc -sum -sub=section main-R1-squeezed.tex > count-new1.txt}

\end{document}